\documentclass[twocolumn,superscriptaddress,pra,reprint,longbibliography]{revtex4-1}
\usepackage{graphicx}
\usepackage{amssymb}
\usepackage{amsmath}
\usepackage{epsfig}
\usepackage{color}
\usepackage{natbib}
\usepackage{mathtools}
\usepackage[colorlinks,linkcolor=blue,anchorcolor=blue,citecolor=blue,urlcolor=blue]{hyperref}
\usepackage{blindtext}
\usepackage{color, soul}

\begin{document}
\setulcolor{red}
\title{Non-Markovianity-assisted high-fidelity Deutsch-Jozsa algorithm in
diamond}
\author{Yang Dong}
\author{Yu Zheng}
\author{Shen Li}
\author{Cong-Cong Li}
\author{Xiang-Dong Chen}
\author{Guang-Can Guo}
\author{Fang-Wen Sun}
\email{fwsun@ustc.edu.cn}

\affiliation{CAS Key Lab of Quantum Information, University of Science and Technology of China, Hefei,
230026, P.R. China}
\affiliation{Synergetic Innovation Center of Quantum Information $\&$ Quantum Physics, University of Science
and Technology of China, Hefei, 230026, P.R. China}\date{\today }

\begin{abstract}
The memory effects in non-Markovian quantum dynamics can induce the revival of quantum coherence which is believed to provide important physical resources for quantum information processing (QIP). However, no real quantum algorithms have been demonstrated with the help of such memory effects. Here, we experimentally implemented a non-Markovianity-assisted high-fidelity refined Deutsch-Jozsa algorithm (RDJA) with a solid spin in diamond. The memory effects can induce pronounced non-monotonic variations in the RDJA results, which were confirmed to follow a non-Markovian quantum process by measuring the non-Markovianity of the spin system. By applying the memory effects as physical resources with the assistance of dynamical decoupling, the probability of success of RDJA was elevated above 97\% in the open quantum system. This study not only demonstrates that the non-Markovianity is an important physical resource but also presents a feasible way to employ this physical resource. It will stimulate the application of the memory effects in non-Markovian quantum dynamics to improve the performance of practical QIP.
\end{abstract}
\maketitle
\vspace{3mm}
\leftline{\textbf{Introduction}}\par
\vspace{3mm}
Based on quantum phenomena, such as superposition,
correlation and entanglement, quantum information processing (QIP) has
provided great advantages over its classical counterpart \cite%
{qip1,qip2,qip3} in efficient algorithms \cite{qa1,qa2}, secure
communication \cite{qc1,qc2}, and high-precision metrology \cite{{qm1,qm2,qm3,JE1,JE2,liu1,liu2,liu3,zhao1}}.

However, quantum superposition and correlation are fragile in an open quantum
system. Notorious decoherence \cite{qdc,WHL1,WHL2}, which is caused by interactions
with noisy environments, is a major hurdle in the realization of
fault-tolerant coherent operation \cite{ftqc,MB1} and scalable quantum
computation. To further expand the implementation of QIP in open quantum
systems, full understanding and control of environmental interactions are required
\cite{qdc,yu1,hanson1,hanson2}. Many techniques have been developed to address this issue, including
decoherence-free subspaces \cite{dfs}, dynamical decoupling (DD) \cite{ddpg,liu1}
and the geometric approach \cite{gqc}. However, actively utilizing the environmental interactions would represent a significant achievement, compared with passively shielding them.

Usually, the interaction of an open quantum system with a noisy environment
exhibits memory-less dynamics with an irreversible loss of quantum
coherence, that can be described by the Born-Markov approximation \cite{nm4}. However,
because of strong system-environment couplings, structured or finite reservoirs,
low temperatures, or large initial system-environment correlations, the
dynamics of an open quantum system may deviate substantially from the
Born-Markov approximation and follow a non-Markovian process \cite{nm4,nm5,nm7,nm8,nmn}. In such a process, the pronounced memory effect,
which is the primary feature of  a non-Markovian environment, can be used to
revive the genuine quantum properties \cite{nm4,nm5,nm7,nm8,nm11,MB2,nmn}, such as
quantum coherence and correlations. Consequently,
improving the performance of QIP by utilizing memory effects as important physical resources in the non-Markovian environment is crucial \cite{nm2,nm3}. However, no
real quantum algorithms have been demonstrated with the help of such memory effects.
Here, we first investigated the memory effects of non-Markovian
environments using the quantum Deutsch-Jozsa algorithm \cite{DJA} with a solid spin
in a diamond nitrogen-vacancy (NV) center. The memory effects were further applied as important physical resources \cite{nm4} to substantially improve
the performance of the quantum algorithm
with the assistance of the DD protection method \cite{ddpg}.

\begin{figure*}[tbp]
\centering
\includegraphics[width=18cm]{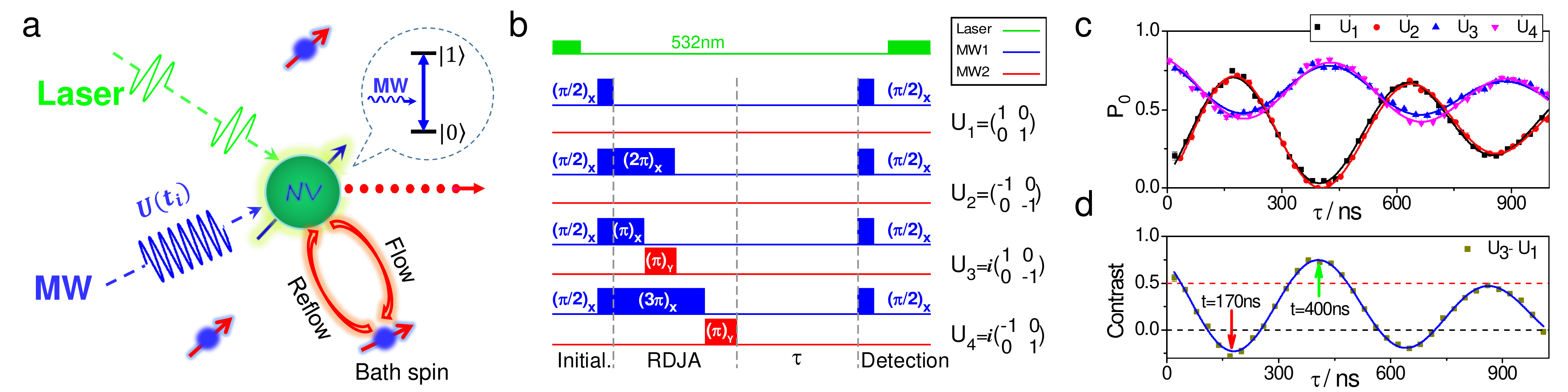}
\caption{(a) The single spin qubit of the NV center was used to study the memory effect of a non-Markovian environment with a bidirectional flow of information between the spin qubit and the
spin bath. (b) Diagram of the MW pulse sequences used to realize the single-qubit
RDJA. (c) The results of the RDJA. The POS for balanced operations (${U_{3}}$ and ${U_{4}}$) is $P_{0}$; whereas for constant operations (${U_{1}}$ and ${U_{2}}$), it is $1-P_{0}$. (d) The contrast between $U_{3}$ and $U_{1}$, corresponding to the POS of the RDJA. When the contrast is less than 50\% as shown by the red dashed line, the advantage of
quantum RDJA is lost completely. In addition, the quantum RDJA completely fails when the
contrast is negative as denoted by the black dashed line.}
\label{fig1}
\end{figure*}

The Deutsch-Jozsa algorithm \cite{rdj1,rdj2,rdj3}, which can be used to determine whether a coin is fair or fake in a single examination step \cite%
{dj,rdj3}, is one of the seminal algorithms used to demonstrate the advantages of
quantum computations. Experimentally, the refined
Deutsch-Jozsa algorithm (RDJA) \cite{rdj1,rdj2,rdj3} was implemented with a
single spin qubit of the NV center to study the memory effect of a non-Markovian environment, with a bidirectional flow of information between the spin qubit and the spin bath, as shown in Fig.\ref{fig1}a. The probability of success (POS) of RDJA presents an unexpected non-monotonic dependence on the delay time of the measurement, that is induced by memory effects of the environment and further
confirmed by the non-Markovianity measurement of the open quantum system \cite%
{nm5,nm8}. Based on a complete understanding of the non-Markovian environment,
we took advantage of the memory effect by applying the DD protection method \cite{ddpg} to significantly enhance the
POS of the RDJA to above $97\%$ in a realistic solid spin
system. This result represents a substantial improvement over previous results \cite{rdj2,rdj4}.
In contrast, the transition from
non-Markovian to Markovian dynamics of the NV center was experimentally
realized with a magnetic field, where the POS of the RDJA
decreases monotonically with the delay time of the measurement, as expected in
the Markovian region. The experimental result clearly confirmed that the performance of a practical quantum algorithm can be improved by incorporating the memory effects from the
non-Markovian environment as important quantum resources.
This experimental result should stimulate the development of the control and application memory
effects of non-Markovian environments for future quantum technologies.

\vspace{3mm}
\leftline{\textbf{Results}}\par
\vspace{3mm}
\leftline{\textbf{Quantum qubit in diamond and RDJA}}\par
\vspace{3mm}
An NV center in diamond with an electron spin $S=1$ was applied for the RDJA implementation at room-temperature. The ${m_{s}}=+1$ and ${m_{s}}=-1$ energy
level degeneracy was lifted by applying a magnetic field along the
symmetry axis of the NV center.

By tuning the microwave (MW) frequency to be resonant with ${m_{s}}=0\leftrightarrow {%
m_{s}}=+1$, these two spin states in the NV center could be encoded as qubit $%
\left\vert 0\right\rangle $ and $\left\vert 1\right\rangle $, respectively
\cite{SI}. As shown in Fig.\ref{fig1},
532 nm laser pulses were used for the initial spin state ($%
\left\vert 0\right\rangle $) preparation and final state readout. The RDJA
can be decomposed into two different unitary transformations: rotation operation
and phase-controlled gate \cite{rdj2,rdj3,rdj4}. The rotation operation can be
realized by a ${(\mathrm{\pi }/2)_{X}}$ pulse resonant with the transition
between $\left\vert 0\right\rangle $ and $\left\vert 1\right\rangle $. Here $%
{\phi _{{X}(Y)}}$ denotes the rotation with angle $\phi $ around the $X(Y)$
axis. The phase gates were realized with ${U}={(-\mathrm{\pi }/2)_{X}}{({%
\phi })_{Y}}{(\mathrm{\pi }/2)_{X}}$, where ${\phi }=0,2\mathrm{\pi }$ for
constant operations (${U_{1}}$ and ${U_{2}}$) and ${\phi }=3\mathrm{\pi },%
\mathrm{\pi }$ for balanced operations (${U_{3}}$ and ${U_{4}}$) in the RDJA. After taking the commutation relations of Pauli matrices, unitary
gates of the RDJA were constructed experimentally, as shown in
Fig.\ref{fig1}b. Finally, another ${(\mathrm{\pi }/2)_{X}}$ was applied
with delay time $\tau$ immediately after MW operations to transfer the relative phase between $\left\vert 0\right\rangle $ and $%
\left\vert 1\right\rangle $ to the spin population for the quantum state measurement.\par

In theory, when balanced operations of the RDJA are applied, the ideal results
for the NV center remain in the bright state $\left\vert 0\right\rangle $. The
POS corresponds to $P_{0}=\mathrm{tr}(\left\vert
0\right\rangle \left\langle 0\right\vert \rho )$, where $\rho $ is the state
after operations. The dark state $\left\vert 1\right\rangle $
occurs for constant operations with the POS $P_{1}=\mathrm{tr}%
(\left\vert 1\right\rangle \left\langle 1\right\vert \rho )=1-P_{0}$. The
results of applying the RDJA to a single NV center are shown in Fig.\ref{fig1}c. The POS of the RDJA is the
contrast between the results of constant and balanced operations, where the
contrast ($P_{0(U_{3})}-P_{0(U_{1})}$) between $U_{3}$ and $U_{1}$ is shown
in Fig.\ref{fig1}d. Clearly, the POS of the
constant (${U_{1}}$ and ${U_{2}}$) operations, balanced (${U_{3}}$ and ${U_{4}}$)
operations, and RDJA show prominent oscillations with the delay
time of the measurement. This result obviously contradicts previous results \cite%
{rdj3,rdj4}. In the Born-Markovian decoherence region \cite{noom}, as the duration of the interaction of the quantum system with the environment increases, the
degeneration of the coherence of the quantum system also increases. Consequently, the
POS should decrease monotonically with the delay time of the
measurement. Moreover, we also observed that quantum RDJA is worse than the
classical method when the contrast is less than $50\%$ and completely fails
when the contrast is negative as shown in Fig.\ref{fig1}d. However,
with an appropriate delay, the maximal POS
is achieved, as shown by the green arrow in Fig.\ref{fig1}d.

\vspace{3mm}
\leftline{\textbf{Non-Markovian effect of noise environment on}}\par
\leftline{\textbf{RDJA}}\par
\begin{figure}[tbp]
\centering
\includegraphics[width=8cm]{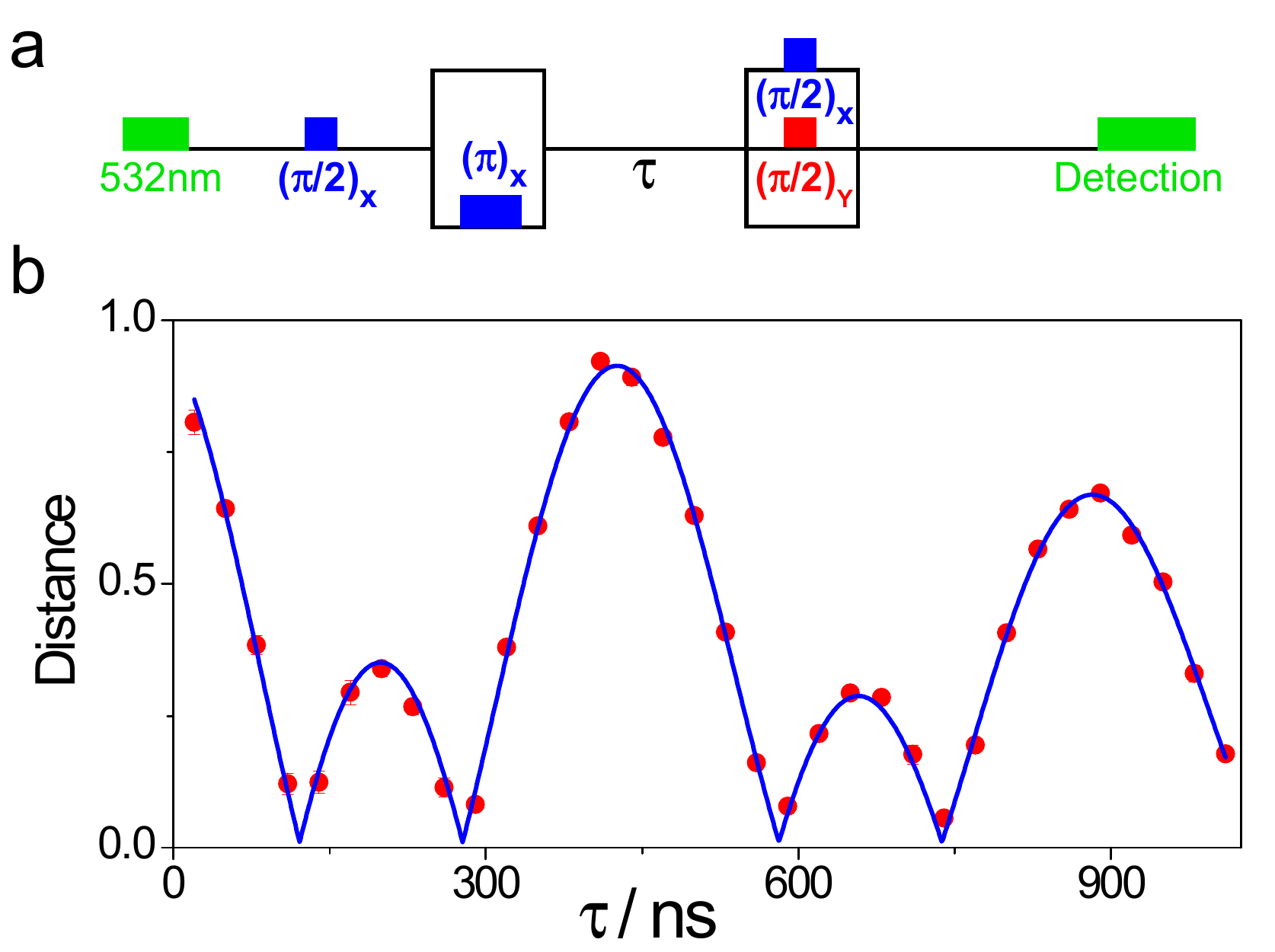}
\caption{(a) Pulse sequences used to characterize the non-Markovianity of the spin
environment. (b) The trace distance of the two optimal states shows the non-Markovianity of the spin environment with revivals. The red dots represent the
experimental results of the trace distance evolution and the blue solid line
denotes the fitting obtained with the supposed model. }
\label{fig2}
\end{figure}

\vspace{3mm}
These nontrivial phenomena deviate mainly from the quantum Born-Markovian process \cite{nm4,nm5,nm7,nm8} and reflect the occurrence of
non-Markovian dynamics. To confirm this supposition and study how
the memory effects of the environment affect the RDJA, we measured the non-Markovianity
of the quantum system by employing the trace distance method \cite{nm5,nm8,nm9},
which is given by
\begin{equation}
N=\mathop {\max }\limits_{{\rho _{1,2}}(0)}\int_{\sigma >0}{\sigma (t,%
{\rho _{1,2}}(0))dt},
\end{equation}%
where $\sigma (t,{\rho _{1,2}}(0))=(\mathrm{d}/\mathrm{d}t)D({\rho _{1}}(t),{%
\rho _{2}}(t))$ is the rate of the change of the trace distance, and $D({\rho _{1}},{%
\rho _{2}})=\left\Vert {\rho _{1}}-{\rho _{2}}\right\Vert /2$, for the two
states ${\rho _{1}}$ and ${\rho _{2}}$. When $N>0$, the interaction process is non-Markovian. To experimentally characterize a non-Markovian quantum dynamics, optimal state pairs \cite{nm10}, $%
\left\vert \circlearrowleft \right\rangle =\left( {\sqrt{2}/2}\right) \left(
{\left\vert 0\right\rangle -i\left\vert 1\right\rangle }\right) $ and $%
\left\vert \circlearrowright \right\rangle =\left( {\sqrt{2}/2}\right)
\left( {\left\vert 0\right\rangle +i\left\vert 1\right\rangle }\right) $,
were experimentally prepared. After the system had interacted with the environment, quantum
state tomography was performed to analyze the dynamics of the quantum
system \cite{SI}, as presented in Fig.\ref{fig2}a. Fig.\ref{fig2}b shows that the
trace distance of the electron spin NV center decreases non-monotonically
with the spin bath interactions. This behavior was the characteristic of the non-Markovianity of the environment.
The measured value of the non-Markovianity is $N = 1.96 \pm 0.14$, which is much larger than $0$ and confirms that this environment is non-Markovian.

This non-Markovian environment of the NV center can be further characterized using a triple-mode spin bath \cite{SI}. The experimental data in Fig.\ref{fig2}b can be fitted with $D({\rho _{1}},{\rho _{2}})=\left\vert (a+b\cos (2\pi \Delta t))\mathrm{e}{^{-t^{2}/T^{2}}}\right\vert$, where $a$ and $b$ relate to the mode density distribution of reservoir,
and $T$ relates to the width of reservoir. We obtained $a=0.105\pm 0.002$, $b=0.218\pm 0.003$, the splitting of the triple-mode reservoir  $\Delta =(2.170\pm
0.005)$ \textrm{MHz} and $T=1382\pm 31$\textrm{ns}. So the width of each
modes was ${724}\pm 16$ \textrm{kHz}. From the type of lineshape which can be explained by coupling with a dark spin $1$ system, the non-Markovian dynamics is from the coupling between the intrinsic $^{14}N$ nuclear spin and the electron spin in NV center.

In fact, if the quantum operations are ideal, the RDJA process is similar
to the preparation of optimal state pairs for the measurement of
non-Markovianity. The oscillation of the trace distance and POS of RDJA implementation is identical, as shown in Fig.\ref{fig1}d
and Fig.\ref{fig2}b by taking $\left\vert
P_{0(U_{3})}-P_{0(U_{1})}\right\vert $. The non-monotonic behavior can be
interpreted as a bidirectional flow of information between the electron spin
of the NV center and the spin bath, which induces the distance between the
two states to decrease or increase. Because the rate of information exchange is on the same order of magnitude as the coherent operation speed, the flow of information begins and tightly accompanies with the RDJA implementation. Therefore, the maximal POS for the RDJA cannot be obtained using a delay of $\tau=0\textrm{ns}$. Indeed, an appropriate delay ($\tau=400 \textrm{ns}$) is required to achieve the maximal POS when the information reflows from the environment, as shown in Fig.\ref{fig1}d.

\begin{figure}[tbp]
\centering
\includegraphics[width=8cm]{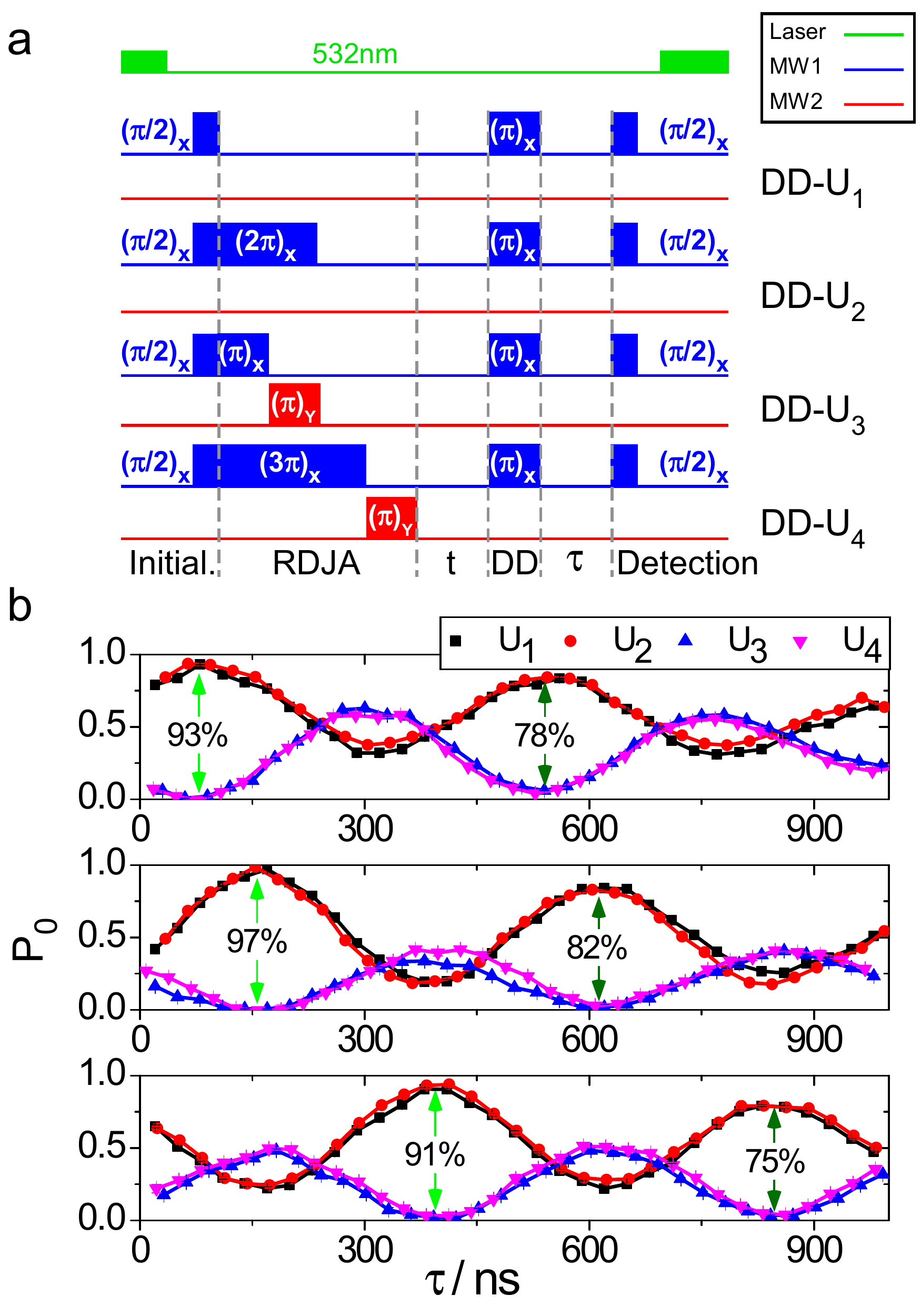}
\caption{(a) Diagram of the 532 nm laser and MW pulse sequence for the RDJA
protected by spin echo sequences. (b) The results of the RDJA protected by
spin echo sequences for $t=76$ ns, $170$
ns, and $400$ ns, with the maximal POS of the RDJA exceed 93\%, 97\%, and 91\%, respectively. The positive and negative echoes are indicated by green arrows.}
\label{fig3}
\end{figure}
\vspace{3mm}
\leftline{\textbf{The performance of DD-protected RDJA under}}\par
\leftline{\textbf{a non-Markovian
environment.}}\par
\vspace{3mm}
The RDJA achieves a POS of approximately $78\%$,
which is relatively high \cite{rdj2,rdj3,rdj4}. However, a recent study revealed
that the memory effects of a non-Markovian environment can be regarded as
important physical resources \cite{nm4,nm11} to improve QIP in an open quantum
system. To further enhance the POS of the RDJA, we utilized the memory effects of the non-Markovian environment, which can be extracted by the DD
method \cite{rdj2,ddpg} to mitigate imperfect operations and
decoherence.

As shown in Fig.\ref{fig3}a, a single ${\mathrm{\pi }}$ pulse between the RDJA and
the measurement was applied to construct the simplest DD sequences: spin echo. Without
complex design sequences \cite{ddpg}, the DD-protected RDJA can be implemented in $%
700$ \textrm{ns} which is mainly limited by the Rabi frequency. Fig.\ref%
{fig3}b shows the results of the implementation of the DD-protected RDJA,
where the positive and negative echoes correspond to the
constant and balanced functions, respectively. Because balanced and constant operations are different in the quantum circuits, as shown in Fig.\ref{fig3}a, the echoes in these operations are expected to appear at different absolute times. Thus we shift each result to facilitate interpretation \cite{rdj2}. Similar to the original
results as shown in Fig.\ref{fig1}c, the POS also oscillates with the delay time of the measurement because of the
non-Markovianity of the environment. However, corrected results can be obtained at the echo delay time, when $t=76, 170,$ and $400$ \textrm{ns}.
When $t=170$ \textrm{ns}, which corresponds
to the failed operation shown in Fig.\ref{fig1}d, we obtained a perfectly corrected result via the assistance of the non-Markovian environment \cite{SI}, as presented in Fig.\ref{fig3}b.

The maximal POS exceeds $97\%$ in the DD-protected RDJA. Moreover, the contrast between the
next local maximum and minimum, which correspond to constant and
balance operations, also exceeds $50\%$ as shown in Fig. \ref{fig3}b. This finding indicates that the DD protection method can enhance the fidelity of
coherent operation and extend the detection region under non-Markovian environment.

\vspace{3mm}
\leftline{\textbf{Transition from a non-Markovian to a Markovian}}\par
\leftline{\textbf{environment.}}\par
\vspace{3mm}
Generally, the non-Markovianity of the present NV center system, which arises from the
structured environments, leads to a non-monotonic dependent relationship
between the POS of the RDJA and the delay time of the measurement. Therefore,
a non-monotonic behavior can be changed by polarizing the spin bath in
diamond \cite{chen1,dnp1,dnp2}.

\begin{figure}[tbp]
\centering
\includegraphics[width=8.7cm]{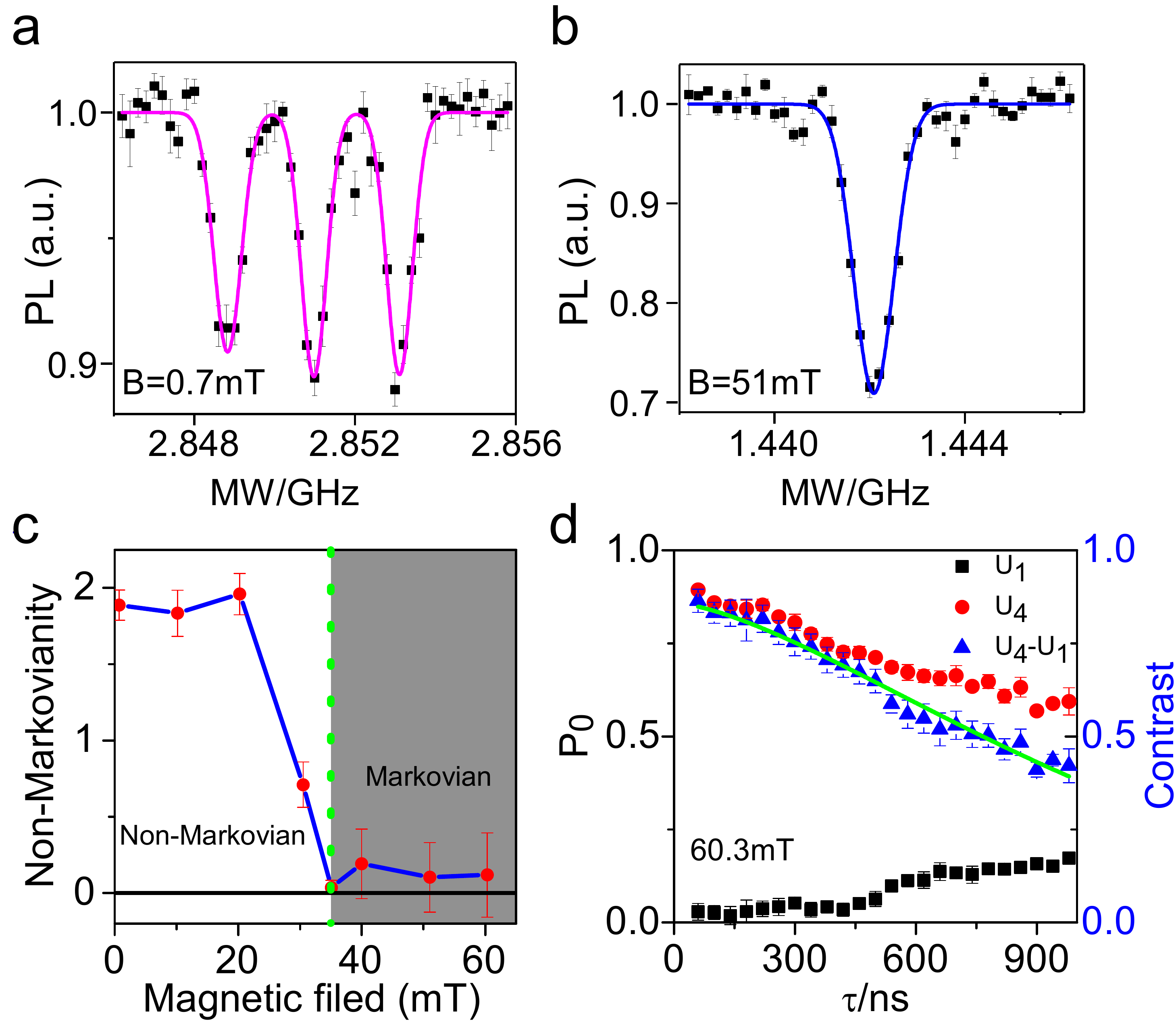}
\caption{(a)-(b) The resonance peaks of a single NV center system under different magnetic field. (c) The transition from the non-Markovian regime to the Markovian regime
at $B=35$ \textrm{mT} (denoted by green dotted line) with the
orientation along the symmetry axis of the NV center. The non-Markovianity measure of Eq.(1)
is obtained by the summation of the differences of the trace distance between the
local minimum and subsequent maximum with a duration of 1$\protect\mu s$. The
black solid line shows $N =0$, which theoretically corresponds to Markovian
dynamics. (d) The results of the RDJA in the Markovian
environment and the pulse sequence shown as Fig.\ref{fig1}b. The contrast (i.e., the POS of the RDJA) decreases monotonically with the delay time of the measurement. }
\label{fig4}
\end{figure}

Here we adopted this strategy \cite{chen1,dnp1,dnp2,dnp3} by making use of a level
anti-crossing in the excited state of the NV center to polarize the
nuclear spins in diamond. By applying magnetic field along the axis of NV center, the tripe-mode spin bath was converted to a single mode system, as shown in Fig.\ref{fig4}a,b. Fig.\ref{fig4}c shows that the non-Markovianity
of spin bath decreases as the magnetic field magnitude increases, while maintaining its orientation along the symmetry axis of the NV center. When the magnetic field exceeded $35$ \textrm{mT}, we realized the transition from
non-Markovian to Markovian dynamics in realistic solid spin system at
room-temperature \cite{SI}. Since such a non-Markovian environment of NV center is caused by coupling with the single intrinsic $^{14}N$ spin, the transition is always same for different single NV centers in diamond. In a Markovian environment, the information re-flow is turned off from spin bath to the electron spin of the NV center. Therefore, the
POS of the RDJA decreases monotonically with the
delay time of the measurement \cite{rdj3,rdj4} as shown in Fig.\ref{fig4}d.
Because of the partial polarization of the spin bath, the maximal POS
increases to $86\%$ compared with the result obtained in the
non-Markovian environment without delay.

\vspace{3mm}
\leftline{\textbf{Discussion}}\par
\vspace{3mm}
We first studied the memory effects of
non-Markovian quantum dynamics on the RDJA with a realistic solid spin system in diamond. The POS of the RDJA exhibits non-monotonic dependence on the delay time of the measurement. Specifically,
we observed that the RDJA would be better than the classical method or fail
completely depending on the delay time of the measurement. The POS
was elevated beyond $97\%$, much higher than
a similar result obtained previously \cite{rdj2,rdj4}, by utilizing the memory effects of the
environment and the DD protection method. Furthermore, we also realized the
transition from non-Markovian to Markovian dynamics in a realistic solid
spin system by applying a magnetic field at room-temperature. Because of the turnoff of the information re-flow from the spin bath to the electron spin of the NV center, the POS of the RDJA decreased monotonically with the delay time of the measurement \cite{rdj3,rdj4}. The memory effects in the non-Markovian process, obtaining the final measurement immediately after the operation is not an optimal strategy when the operation speed is comparable to the rate of information exchange between the system and the environment. However, using an appropriate delay time recovers the result. This study also demonstrated that the memory-effect-based non-Markovianity
can be used as an important physical resource. Thus extracting and applying this resource for quantum information techniques is feasible. This finding will
stimulate the application of these memory effects in non-Markovian quantum dynamics to improve
the performance of practical QIP.\par
Furthermore, the spin state of the NV center in diamond was shown to be an excellent test platform for the study and application of non-Markovian dynamics
beyond elaborate engineering systems \cite{nm8,nm9}. In the future, the electron spin of the NV center could be used to address some fundamental physical questions \cite{nm4} such as the mathematical structure of the
geometric space of non-Markovian quantum dynamical maps, the relevance of
non-Markovianity in the study of the border between classical and quantum
aspects of nature, and the use of non-Markovian quantum probes to detect nonlocal initial
correlations in composite environments. Ultimately, with a deep understanding of
and perfect control over the environment, the performance of QIP in solid-state system can be further improved.

\vspace{3mm}
\leftline{\textbf{Methods}}\par
\vspace{3mm}
We used a room-temperature home-built confocal microscopy, with a dry objective lens (N.A. $= 0.9$), to address single NV center in a type-IIa, single-crystal synthetic diamond sample
(Element Six). The abundance of $^{13}$C was at the nature level of 1\%. The NV centers were produced by nitrogen ions implantation and the energy of the implantation was 30 $\textrm{keV}$.
The dosage was ${10^{11}}$$/{\textrm{c{m}}^2}$ and the estimated average depth of NV was $20$\textrm{nm}. For NV center, the environment was made with dark spin bath $%
^{13}$C and $^{14}$N. In our work, the former was come from diamond lattice
and was the source of slow dephasing noise. And the latter was come from
nitrogen implantation, which caused the non-Markovianity of environment for
NV center. The NV center, mounted on a three-axis, closed-loop piezoelectric stage for sub-micrometre-resolution scanning, was illuminated by a $532$ \textrm{nm} diode laser. Fluorescence photons (wavelength ranging from 647
nm to 800 nm) were collected into a fiber and detected using the
single-photon counting module, with a counting rate of
$130$ \textrm{kHz} and a signal-to-noise ratio of $200:1$. We verified single photon emission from the NV center by measuring the photon correlation function. An impedance-matched gold coplanar waveguide (CPW), deposited on the bulk diamond, was used for delivery of MW to the NV center. The optical and MW
pulse sequences were synchronized by a multichannel pulse generator (Spincore, PBESR-PRO-300). \par

\vspace{3mm}
\leftline{\textbf{Acknowledgements}}\par

This work is supported by The National Key Research and Development Program of China (No. 2017YFA0304504), the National Natural Science
Foundation of China (Nos. 11374290, 61522508, 91536219, and 11504363).

%

\clearpage
\newpage
\leftline{\textbf{Supplementary Information}}

\section{EXPERIMENTAL SETUP}
\begin{figure}[tbp]
\centering
\includegraphics[width=8.7cm]{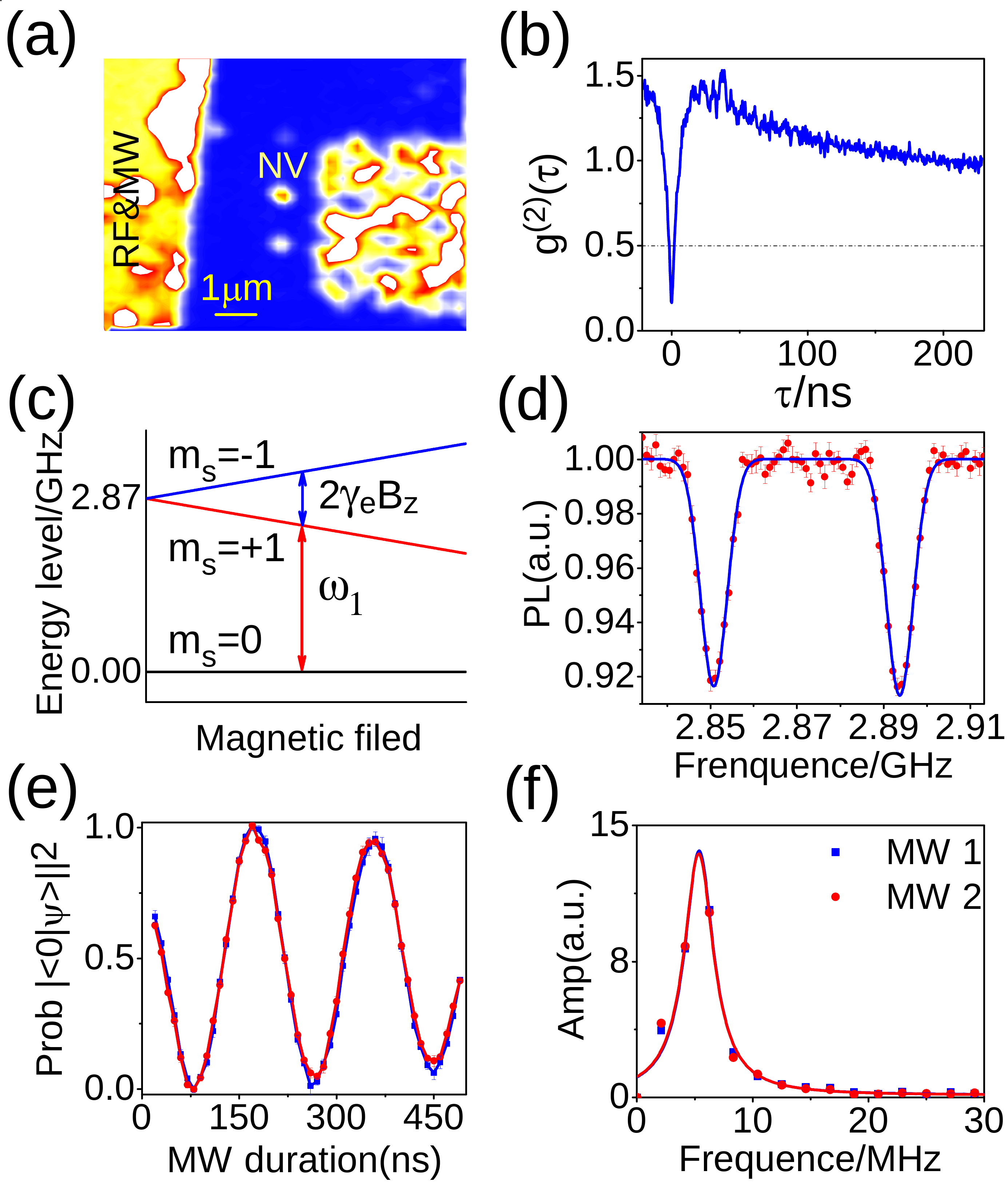}
\caption{(a) Confocal image of the NV center used in the experiment. A
coplanar waveguide antenna is deposited to deliver microwave pulses to the
NV center. (b) Fluorescence correlation function. (c) Energy level diagram
of the electronic ground state under magnetic field. (d) ODMR spectra for
the single NV center. The transition frequencies for ${m_s} = 0
\leftrightarrow {m_s} = +1$ and ${m_s} = 0 \leftrightarrow {m_s} = - 1$ are
2.8507 GHz and 2.8915 GHz respectively. (e) Rabi oscillations with two MW
channels at frequency 2.8507 GHz. (f) FFT spectra of (e) show a 5.37MHz Rabi
oscillations}
\label{SFIG1}
\end{figure}

Fig. \ref{SFIG1}(a) shows the image of NV center detected by a home-built confocal
microscopy at room-temperature. We verified single photon emission from the NV center by measuring the
photon correlation function ${g^{2}}(\tau )$ as shown in Fig. \ref{SFIG1}(b). And ${%
g^{2}}(0)<1/2$ indicates a single NV center.

With the secular approximation, the effective Hamiltonian for the qubit system
reads \cite{ddpg,ddpq1}
\begin{equation}
H={\omega _{e}}{S_{z}}+{S_{z}}\sum\nolimits_{j}{{A_{j}}I_{z}^{j}}%
+\sum\nolimits_{j}{{\omega _{I}}I_{z}^{j}}+{H_{dip}},
\end{equation}%
where ${\omega _{e}}$(${\omega _{I}}$) is the electron (nuclear) spin
resonance frequency, ${S_{z}}$(${I_{z}^{j}}$) denotes electron (nuclear)
operator, ${A_{j}}$ is the hyperfine coupling between the electron and the $%
j $th nuclear spin, and ${H_{dip}}$ denotes the interaction within the
nuclear spin bath. Electron spin levels ${m_{s}}=\pm 1$ are separated from ${%
m_{s}}=0 $ by a zero-field splitting $D\approx 2.87\mathrm{GHz}$ \cite{chen2}%
. The ${m_{s}}=+1$ and ${m_{s}}=-1$ energy level degeneracy is lifted by a
static magnetic field along the NV symmetry axis as shown in Fig. \ref{SFIG1}(c). The
optically detected magnetic resonance (ODMR) spectra of NV center is shown
in Fig. \ref{SFIG1}(d). By tuning microwave (MW) frequency resonance with ${m_{s}}%
=0\leftrightarrow {m_{s}}=+1$, the NV center can be treated as a pseudo-spin-1/2
system \cite{hanson2}.

In the experiment, we used two MWs (MW$1$ and MW$2$) for rotation operations
around $X$ and $Y$ axes, respectively. Before the implementation of the RDJA
\cite{rdj1,rdj2,rdj3}, we first measured the frequencies of the Rabi
oscillations with MW$1$ and MW$2$ as shown in Fig. \ref{SFIG1}(e) and made sure the
same power of them. In order to analyze Rabi frequencies in these two MW
channels, fast Fourier transform (FFT) spectra of the Rabi oscillations were
carried out as shown in Fig. \ref{SFIG1}(f). We used $\pi $ pluses with the length of
93ns for these two MW channels. The relative phase between MW$1$ and MW$2$
was $\mathrm{\pi }/2$ and was verified by spin-locking experiment.


\section{SIGNAL INTERPRETATION}
In the experiment, the observed fluorescence signal, which is related to the
population distributions among ${m_s} = 0 $ and ${m_s} = +1$ states of NV
center, can be converted to the probability of success of quantum coherent
operation by linear transformation. Specially, we firstly initialized the
system into ${m_s} = 0 $ state by a laser pulse and then changed it into ${m_s} = +1$ state by MW operation and measured their photon counts (denoted
by $C_{\max}$, $C_{\min}$). In order to beat the fluctuation of photon
counting, we repeated the experimental cycle at least ${10^6}$ times. So the
relative population of the ${m_s} = 0 $ state for an unknown state can be
expressed as
\begin{equation}
{P_0} = \frac{{C - {C_{\min }}}}{{{C_{\max }} - {C_{\min }}}},
\end{equation}%
where C is the measured photon count under the same experimental condition.
Therefore, the probability of successful quantum operation can be expressed
as the function of fluorescence intensity of NV center.

\section{THE REFINED DEUTSH-JOZSA ALGORITHM}
The main idea of the refined Deutsh-Jozsa algorithm \cite{rdj1} with single qubit is used to determine whether a coin is fair or fake just by one step. However, in classical world, we have to check both side of coin. Here, we give detailed instructions of it.\par
Step 1: initializing quantum system to ground state $\left| 0 \right\rangle $£»\par
Step 2: implementing Hadamard gate on the qubit. The state will be at $\left| 0 \right\rangle  + \left| 1 \right\rangle $;\par
Step 3: operating a quantum black box on the qubit. If the coin is fake, we get $\left| 0 \right\rangle  - \left| 1 \right\rangle $. If the coin is fair, we get $\left| 0 \right\rangle  + \left| 1 \right\rangle $. In the experiment, the black box of fake coin corresponds to constant operation and the black box of fair coin corresponds to balance operation;\par
Step 4: implementing Hadamard gate again on qubit. If the coin is fake, we get $\left| 0 \right\rangle $. If the coin is fair, we get $\left| 1 \right\rangle $;\par
Step 5: readout.\par
The realization of quantum black box is the core of the quantum algorithm. In RDJA, the quantum core is a phase gate. In our work, we make use of a single qubit phase gate (${e^{ - i\theta {\sigma _z}}} = {e^{i\pi /2{\sigma _x}}}{e^{ - i\theta {\sigma _y}}}{e^{ - i\pi /2{\sigma _x}}}$) to realize it. A third auxiliary level of NV center can also be applied to make the gate \cite{rdj2}. However, the single qubit phase gate can be carried out feasibly without additional microwave driving \cite{rdj2}.\par
\begin{figure}[tbp]
\centering
\includegraphics[width=8cm]{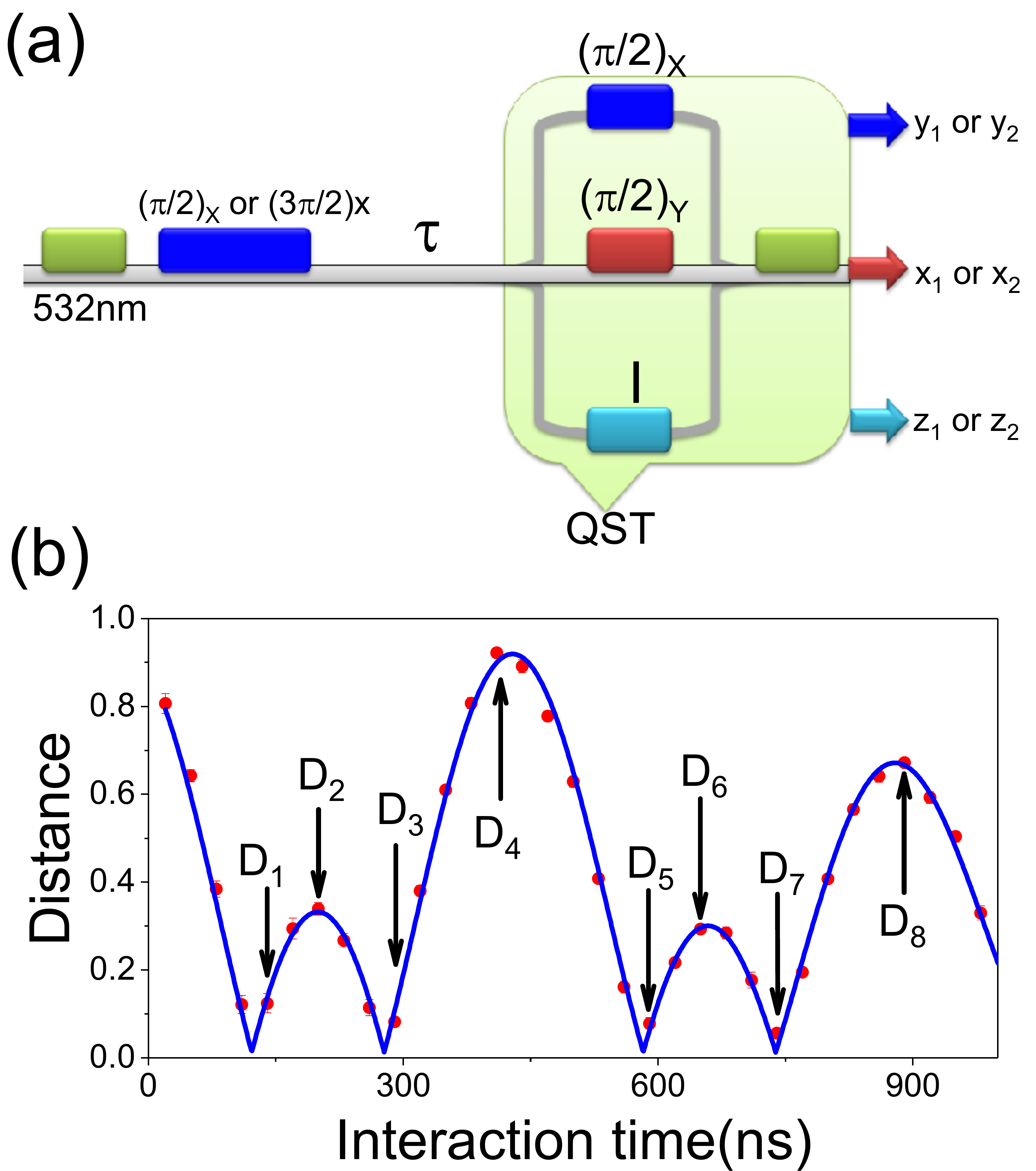}
\caption{(a) Characterisation of the non-Markovianity of the spin environment. It represents: initializing the system to ground state, preparing optimal state pairs to probe the non-Markovianity of the environment, and measuring the density matrix of quantum system by single qubit QST. (b) The trace distance
of the two optimal states under noise environment.}
\label{SFIG2}
\end{figure}

\section{TRIPLE MODE OF SPIN BATH}
In the main text, we witnessed the strong non-Markovian environment of the
quantum system by the trace distance method. A nonzero value for this method means that there is an initial state pair which the trace distance increases over a certain time interval. It can be interpreted as a flow of information from the environment back to the open system, implying the presence of memory effects. A pair of states, ${\rho _1},{\rho _2}$ are said to be an optimal state pair\cite{nm9} if the maximum of $N$ is attained. For NV center in the diamond, the environment is given by the nuclear spins coupling to the electron spin of NV center via hyperfine interaction\cite{hanson2}. Due to energy mismatching between NV center and spin bath, the spin flip-flop between them is usually prohibited. Hence, the energy conservation process such as pure dephasing is the main characteristic of NV center. And this process has direct effect on coherent superposition states, which stay in equatorial plane of Bloch sphere. So according to optimal state pair criterion \cite{nm9}, the optimal state pair is chosen in the present work.

After carrying out quantum
state tomography (QST) as shown in Fig. \ref{SFIG2}(a), we can get the density matrix of the system:
\begin{eqnarray}
{\rho _{1}} &=&\frac{1}{2}\left( {I+{x_{1}}{\sigma _{x}}+{y_{1}}{\sigma _{y}}%
+{z_{1}}{\sigma _{z}}}\right) \text{,} \\
{\rho _{2}} &=&\frac{1}{2}\left( {I+{x_{2}}{\sigma _{x}}+{y_{2}}{\sigma _{y}}%
+{z_{2}}{\sigma _{z}}}\right) \text{,}
\end{eqnarray}%
where all the components of density matrix can be obtained by normalizing
fluorescence signal. Then we can calculate trace distance with formula
\begin{equation}
D({\rho _{1}},{\rho _{2}})=\frac{{\left\Vert {{\rho _{1}}-{\rho _{2}}}%
\right\Vert }}{2},
\end{equation}%
and this result is shown in Fig. \ref{SFIG2}(b). In experiment, the non-markovianity of the quantum system is calculated by \cite{nm8}
\begin{equation}
N = {D_2} - {D_1} + {D_4} - {D_3} + {D_6} - {D_5} + {D_8} - {D_7}.
\end{equation}\par
\begin{figure}[tbp]
\centering
\includegraphics[width=8.5cm]{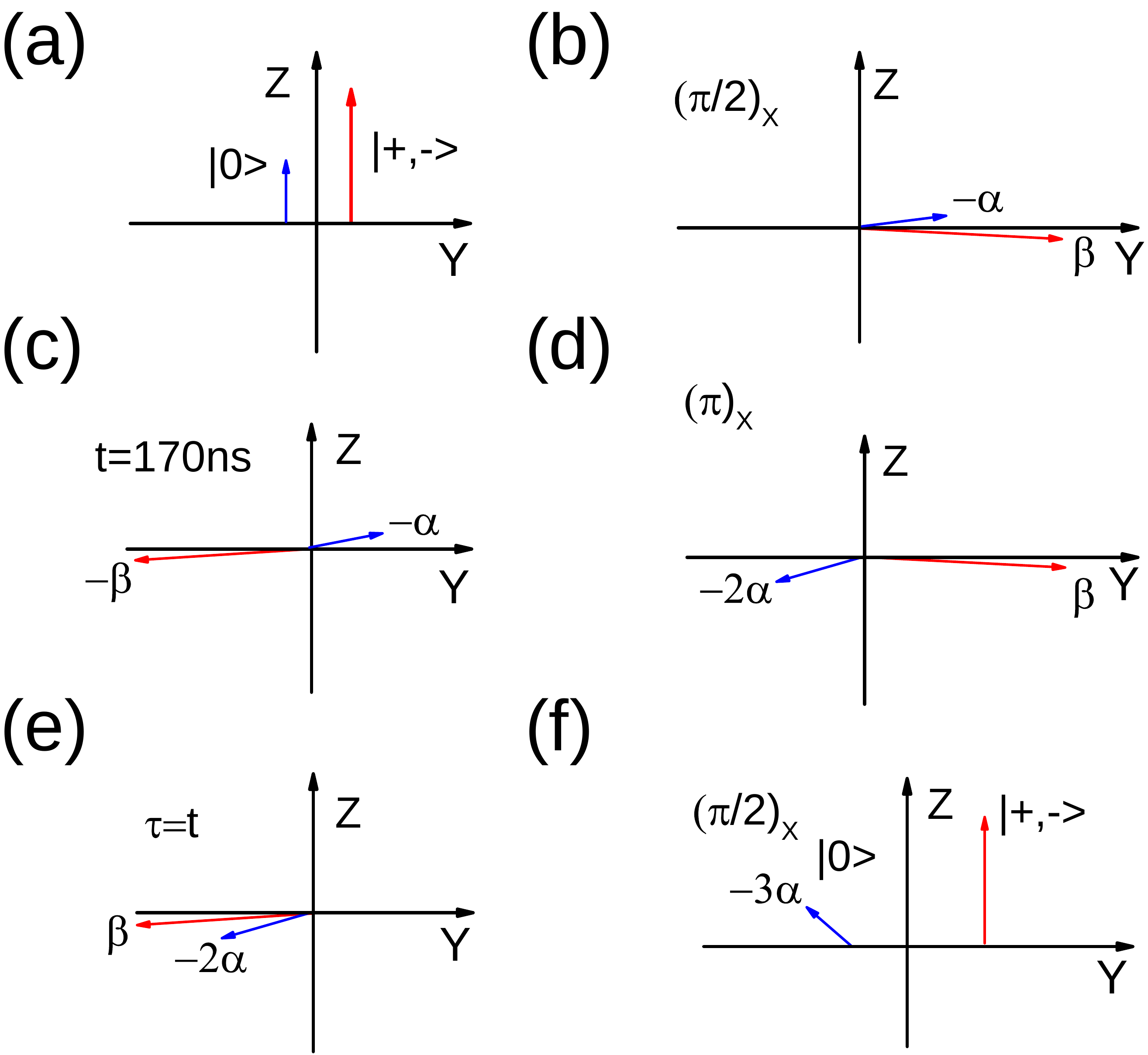}
\caption{Quantum state in the Y-Z plane of the Bloch sphere. 0, +/- denote the tripe-mode spin bath respectively. (a) Initializing the system into ground state. (b) Implementing ${{\text{(}}\pi {\text{/2}})_X}$ gate on initial state. The resonance mode (denotes with 0) rotates slowly than the other two detuned modes. (c) Interacting with spin bath for $t = 170$ ns. (d) Applying imperfect DD sequence. (e) Interacting with spin bath for $\tau  = t = 170$ ns again. (f) Implementing ${{\text{(}}\pi {\text{/2}})_X}$ gate again and reading it out.}
\label{SFIG3}
\end{figure}
For NV center in diamond, the environment is composed of dark spin bath $^{13}$C and $^{14}$N and can be described by $\Delta {S_{z}}{I_{z}}$\cite{hanson2}. Here ${I_{z}}$ is the $z$--component of a dark nuclear spin, and $\Delta $ is the splitting of the triple mode reservoir. In the experiment, MW was resonant
with the center frequency of ODMR and the evolution of trace distance of two optimal states can be fitted with
\begin{equation}
D({\rho _{1}},{\rho _{2}})=\left\vert (a+b\cos (2\pi \Delta t))\mathrm{e}{%
^{-t^{2}/T^{2}}}\right\vert ,
\end{equation}%
where $a$ and $b$ relate to the mode density distribution of reservoir,
and $T$ relates to the width of reservoir. The type of lineshape can be explained by coupling with a dark spin-1 system instead of $1/2$ \cite{nm8}. By fitting the experimental
data, we obtained $a=0.105\pm 0.002$ , $b=0.218\pm 0.003$, $\Delta =(2.170\pm
0.005)$ \textrm{MHz} and $T=1382\pm 31$\textrm{ns}. So the width of each
modes was ${724}\pm 16$ \textrm{kHz}. This result is agree with structure of resonance peaks as shown in Fig. 4(a).

The frequency spectrum structure can be characterized by a triple-mode,
\begin{equation}
S(\omega ) = {A_1}{\operatorname{e} ^{ - 2{{\left( {\frac{{\omega_e  - \Delta }}{w}} \right)}^2}}} + {A_2}{\operatorname{e} ^{ - 2{{\left( {\frac{\omega_e }{w}} \right)}^2}}} + {A_3}{\operatorname{e} ^{ - 2{{\left( {\frac{{\omega_e  + \Delta }}{w}} \right)}^2}}} \text{,}
\end{equation}
where ${\omega_e}$ is the electron spin resonance frequency. The effective Hamilton for NV center under MW driving reads
\begin{equation}
H = {\omega _e}{\sigma _z}/2 + \Omega \cos ({\omega _e}t){\sigma _x} + f(t){\sigma _z}/2 ,
\end{equation}
and we define $g({t_1} - {t_2}) = \left\langle {f({t_1})f({t_2})} \right\rangle. $ So $S(\omega )$ corresponds to the Fourier transform of $g(t)$. With this model, we can get the physical phenomenal formula Eq. (7).
\par

Due to the interaction between the spin qubit and the tripe-mode spin bath, the quantum gate, which was based on the Rabi oscillation, was imperfect and showed an average result. However, DD method can be applied to the modify the imperfect operation. After we prepared imperfect state which is denoted in the Y-Z plane of Bloch sphere, as shown in Fig. \ref{SFIG3}(b), we let it interact with the spin bath with a duration time $t \approx 170$ ns, corresponding to complete failure of the RDJA for original sequence as show in Fig. 1(c) of main text. After applying DD operation, we got a state as shown in Fig. \ref{SFIG2}(d). After the system interacted with spin bath for $\tau  = t$ again as shown in Fig. \ref{SFIG3}(e), we implemented ${{\text{(}}\pi {\text{/2}})_X}$ gate and read the state out as shown in Fig. \ref{SFIG3}(f). From quantum state dynamic evolution, we can see only when $t \approx (170 + n{\Delta ^{ - 1}})$ ns (${\text{n  =  0, 1, 2 }} \cdots $) the imperfect operation caused by spin bath can be modified by itself perfectly except the middle modes of spin bath which was resonant with MW. In brief, the DD method filters the slowly noise and non-Markovian environment modifies imperfect operations.

\begin{figure}[tbp]
\centering
\includegraphics[width=8.7cm]{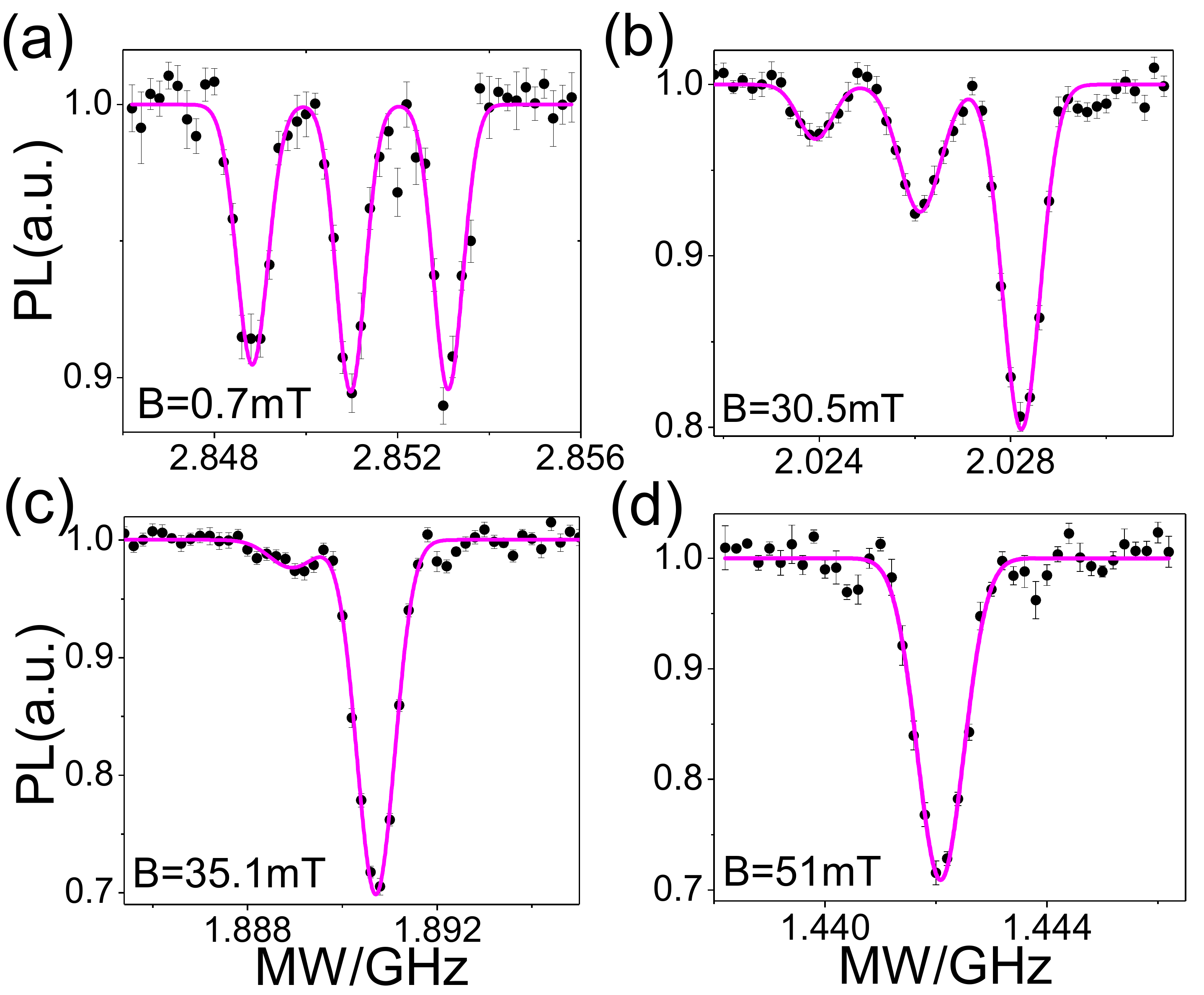}
\caption{(a)-(d) The frequency spectrum of the initial state under different magnetic field.}
\label{SFIG4}
\end{figure}

Fig. \ref{SFIG4}(a)-(d) show the spin bath for single NV center from a three-mode to a single mode structure by applying magnetic field along the axis of NV center. With this method, we can control the open system dynamics to observe the transition between non-Markovian and Markovian quantum mapping.

\end{document}